# ceRNAs 网络的构建揭示了前列腺癌的预后特征

陈梦婕 [1]

[1]湖南大学生命医学交叉研究院，长沙 410000

chenmengjie@hnu.edu.cn

【摘要】

转录后调控异常是诱导肿瘤发生的主要原因之一。近年来，越来越多的证据显示竞争性内源性 RNAs (ceRNAs)可以通过竞争彼此共有的 microRNA(miRNA) 介导转录后调控机制，从而参与肿瘤发生发展各阶段。在本研究中，我们对 ceRNAs 在前列腺癌(PCa)中的调控机制和功能作用进行了全面的研究，构建了一个对患者预后具有潜在价值的 ceRNAs 网络，并将其作为前列腺癌的治疗靶点进行评估。

【关键词】

前列腺癌；ceRNA 调控网络；microRNA

【abstract】

The dysregulation of transcripts is characterized as one of the main mechanisms in tumor pathogenesis. The recent discovery developed a new hypothesis, competitive endogenous RNAs (ceRNAs), which could regulate other RNA transcripts via competing for their shared miRNAs. The interaction of elements in ceRNAs network was involved in a large range of biological reactions and facilitate to cancer progression. In this study, we performed a comprehensive investigation on the regulatory mechanisms and functional roles of ceRNAs in prostate cancer (PCa) and constructed a ceRNAs network which could possess potential value in patient prognosis and be evaluated as therapeutic targets for PCa.

【keyword】

Prostate cancer；ceRNA network；microRNA

【引言】

前列腺癌是男性患者中最常诊断的癌症，也是全球男性癌症死亡的第二大原因。截至 2022 年，约有 268490 例新癌症病例和 34500 例死亡[1]。前列腺是一种外分泌腺，由基底细胞(basal-like cells)和管腔上皮细胞(luminal-like cells)，以及少量神经内分泌细胞(neuroendocrine prostate cancer, NEPC)所组成。由正常前列腺癌变而来的早期前列腺癌阶段称为原发性前列腺癌(primary prostate cancer, pri-PCa)。临床上，大多数低级 pri-PCa 可以通过根治性前列腺切除术或者放疗等方式进行治疗，且预后良好。而局部晚期的 pri-PCa 和转移性前列腺癌主要采用雄激素剥夺治疗(Androgen deprivation therapy, ADT)，例如，促黄体激素释放激素(LHRH)类似物可通过抑制睾丸分泌雄激素发挥作用。但最终大部分患者体内肿瘤会复发并进一步发展为恶性程度更高的去势抵抗性前列腺癌（Castration-resistant prostate cancer, CRPC）。目前，针对 CRPC 的治疗为新一代抗雄激素治疗药物有恩杂鲁胺和阿比特龙，它们是雄激素受体拮抗剂，可阻断雄激素结合雄激素受体并防止配体 - 受体复合物的核移位和共激活剂募集。但不幸的是，接受新一代抗雄激素治疗的患者体内肿瘤会复发并进一步恶化为神经内分泌前列腺癌（Neuroendocrine prostate cancer, NEPC），此阶段几乎无药可医[2,3]。因此，有必要发现 PCa 发生发展的机制，并识别鉴定出新的生物标志物。

随着高通量技术的进步，新型生物标志物的检测将有更广阔的空间。其中，竞争性内源性 RNAs (ceRNAs)网络假说是机体在转录后水平上调控 PCa 发展过程中基因表达失调的重要机制之一。ceRNAs 网络由 ceRNAs 以及它们竞争结合共同的 microRNAs(miRNAs)所组成。这些 ceRNAs 包括蛋白质编码基因(mRNAs)和长链非编码 RNAs（long non-coding RNAs, lncRNAs)，而 miRNAs 是一类估计有 22 个核苷酸长度的非编码 RNA，可通过结合靶 RNA 的互补序列抑制靶基因的表达，从而参与 PCa 的发展进程。基于 miRNAs 与靶 RNA 的相互作用，ceRNAs 网络调控假说应运而生，即 ceRNAs 因竞争结合共有的 miRNAs 而相互交叉调节，从而抑制 miRNAs 介导的靶基因沉默[4-6]。越来越多的证据表明，ceRNA 网络失衡可能在肿瘤发生中起重要作用[5,7]。例如，在 PCa 中，PTEN 的 ceRNAs(VAPA 和 CNOT6L)可以通过吸附 PTEN 靶向的 miRNAs 以解除 miRNAs 对 PTEN 的负调控，从而抑制 PCa 的发展[8]；lncRNA UCA1 通过与 MYO6 竞争 miR-143 的组合来促进 PCa 的进展[9]；SP1 介导的 lncRNA SNHG4 上调可能通过 PCa 中的 SNHG4/miR-377/ZIC5 促进肿瘤发生[10]。然而，上述研究发现都只揭示了 ceRNAs 调控 PCa 发展网络的冰山一角，而在本课题中，我们基于差异表达基因

(Differentially expressed genes，DEGs)和预测 miRNA 结合位点的五个数据库（targetscan、starbase、miRTarBase miRcode 和 miRDB）建立了一个 ceRNA 调控网络。因此，本研究可从分子机制的角度来更好地理解 PCa 的发生发展，这有助于识别前列腺癌中未被鉴定的预后标志物，并帮助制定出潜在的靶向治疗策略。

# 【结果】

## 1　ceRNA 网络构建的研究框架

研究流程图如图 1 所示。首先，我们分别从 TCGA 数据库中获得正常前列腺组织和 pri-PCa 组织的转录组数据以及 Dan Robinson 的文章中获得的 CRPC 组织的转录组数据[11]，然后通过对比分析筛选出差异 mRNAs(Differential expression, DE mRNAs)和差异 lncRNAs(Differential expression, DE lncRNAs)。接下来，我们对所选 DEGs 进行了功能标注，并基于 ceRNAs 以共表达网络构建的方式绘制了关联图。另外，我们进行了 unique-cox 和 LASSO 回归分析，而后以此筛选出可作为独立危险因素且具有潜在预后价值的 DEGs，并进行了生存期、危险比和 ROC 曲线分析。最后，我们进行了数据集验证、重新分组分析和一系列湿实验去验证我们在 PCa 中构建的 ceRNAs 调控网络(图 1)。

## 2　各阶段 PCa 中 DEGs 的综合分析

为了探究 DE mRNA 在前列腺癌发展过程中的作用，我们首先获得了 333 个 pri-PCa 组织和 52 个正常前列腺组织，以及 98 个 CRPC 组织。聚类分析的结果显示，CRPC 组织能够成功地从其余两组分离出来，然而组成成分分析显示 pri-PCa 和正常前列腺组织无法完全分离(图 2A)。除去实验数据平台均一化的问题外，该结果是符合逻辑的，因为在前列腺癌恶化的过程中，恶性程度较高的 CRPC 组织与正常前列腺组织的表达差异理应比恶性程度较低的 pri-PCa 更为明显。同样，在剔除表达量较低（basemean<5）的基因后，我们最终观察到了正常组织和 pri-PCa 中存在着 4292 个 DEGs，其中包括 2731 个 DE mRNAs 和 1002 个 DE lncRNAs，而在 pri-PCa 和 CRPC 中鉴定出了 9951 个 DEGs，包含 6040 个 DE mRNAs 和 2061 个 DE lncRNAs（图 2B）。

## 3　PCa 中 ceRNAs 调控网络的构建

基于上述筛选得到的 DEGs，我们结合 miRTarBase, starbase, miRDB, TargetScan 四个数据库分析预测出了候选的 DE ceRNAs 以及与它们有潜在结合能力的 miRNAs。为了确保结果的可靠性，在筛选 ceRNAs 组成成分的过程中设置了以下两个条件：（1）候选 ceRNAs 和对应 miRNAs 的结合关系需要被基于实验数据建立的 miRTarBase 数据库预测到；或者二者的结合关系应被至少两个除 miRTarBase 外的数据库预测到（2）miRNAs 在前列腺癌发展阶段的表达具有差异性（图 3A）。只有满足上述条件的其中一条，才可以被初步选入 ceRNAs 调控网络之中。为了进一步缩小 ceRNAs 网络成分的范围，我们只选择了存在潜在调控关系的候选 ceRNAs 和对应 miRNAs（其中 ceRNAs 和 miRNAs 的调控相关系数的绝对值要大于 0.2，ceRNAs 之间的调控关系要大于 0.3），结果获得了 273109 成对的 ceRNAs，包含了 298 个 miRNAs 和 2534 个 mRNAs(图 3B)。

## 4 miR-375 抑制前列腺癌的发展

通过比较 ceRNAs 网络中的 miRNAs 在 pri-PCa 组织与正常组织的表达，我们首先发现 miR-375 在肿瘤组织中表达上调。然而，miR-375 在各种癌症中被报道为肿瘤抑制因子[12-15]。为了进一步研究 miR-375 对 PCa 进展的影响，用 miR-375 mimics（实验组）或寡核苷酸(阴性对照组)转染 PCa 各个细胞系 48 小时，然后用于表型分析。首先发现，与阴性对照组相比，实验组中 miR-375 水平显著提升千倍左右（图 4A）。而通过克隆形成实验以及测量细胞过表达后细胞增殖数量随时间的变化，我们发现相比于对照组，在四个细胞系中过表达 miR-375 可以显著抑制细胞增殖和集落形成能力（图 4B&C）。值得注意的是，在培养细胞的过程中，我们发现 miR-375 过表达组的细胞状态较差，并出现大量死亡（图 4D）。因此，接下来我们用 Annexin V FITC/PI 凋亡检测试剂盒检测各 PCa 细胞系的凋亡情况，结果证明了相比于对照，miR-375 过表达组的 DU145 和 PC3 可以显著促进 PCa 细胞凋亡图 4E）。肿瘤细胞成球实验显示，与阴性对照细胞相比，miR-375 过表达的干性受到抑制（图 4F）。另外，在侵袭实验中，我们根据图片灰度值大小对各组肿瘤细胞能力进行了定量，结果发现 miR-375 能够有效抑制 PCa 的迁移和侵袭能力。细胞划痕进一步证实，miR-375 确实在 PCa 中发挥抑制迁移的功能，且与阴性对照相比，差异有统计意义（图 4G）。接下来，我们用能够封闭 miR-375 功能的 miR-375 抑制剂感染细胞，却没有在克隆形成实验，成球实验以及划痕实验中观察到 PCa 细胞有显著的表型变化（图 5A-F）。但是，当我们同时转染 miR-375 mimics 和抑制剂时，发现 miR-375 抑制剂能够显著恢复 miR-375 引起的抑制前列腺增殖的效

果，尤其是对细胞增殖、凋亡和状态的影响（图 5G-K）。总的来说，这些结果表明 miR-375 可能在前列腺癌发生发展的过程中发挥着抑制作用。

为了进一步确定 miR-375 在体内前列腺癌中的治疗潜力，我们使用异种移植瘤模型来评估 miR-375 的肿瘤抑制能力。值得注意的是，由于 miR-375 抑制剂在细胞中产生的表型效果并不明显，因此，这里我们只将 NC 和 miR-375 过表达的 PC3 细胞(1x106 cells/50ul)皮下植入 BALC/c 裸鼠，而后每隔三天记录肿瘤大并于大约三个月后从小鼠体内取出肿瘤拍照并测量重量（图 6A）。结果表明，与 NC 组相比，miR-375 过表达可导致 PC3 转移瘤大小(p<0.0001)和重量(p=0.0037)显著降低。值得注意的是，miR-375 过表达组肿瘤发生率显著低于 NC 组（图 6B&C）。此外，我们将具有转移能力的 PC3M 细胞(1x106 细胞/100ul)注入 BALC/c 裸鼠构建肺转移模型。结果发现，与 NC 组相比，miR-375 过表达会降低 PC3M 移植瘤的增殖能力（p=0.4424），并且通过肺转移实验，我们证明了 miR-375 能够损害 PC3M 移植瘤的转移能力(p=0.0311)。同样，miR-375过表达组肿瘤发生率也受到了一定程度的抑制（图 6D-F）。综上所述，miR-375不仅可以在体外抑制细胞的生长，并且可以在体内显著抑制肿瘤生长。因此，miR-375 可能具有潜在的治疗价值。

## 5 RNA-seq 分析揭示了 miR-375 调控的途径

在确定 miR-375 在 PCa 发生发展的过程中发挥着重要的调控作用后，我们想进一步从机制的角度深入理解 miR-375 引起的抑癌效果。为了筛选出 miR-375 的新靶点，我们先用 miR-375 mimics 和 NC 寡核苷酸链转染 PC3 和 LNCap 细胞，再利用 RNA-Seq 技术检测了其中全基因组的表达水平并进行了系统地分析（图 7A）。通过对 RNA-Seq 的分析，我们在 LNCap 约 20,000 个已定位的基因中，鉴定出了 643 个 DEGs，其中包含 295 个上调基因和 348 个下调基因。Metscape 显示，这些 DEGs 与多个癌症相关信号通路相关，包括 E2F 通路、DNA 修复、微管细胞骨架和胃癌网络（图 7B）。此外，GSEA 显示 miR-375 的过表达对 PCa 发生发展有抑制作用，尤其是对 MYC 信号通路有一定的抑制作用，而且我们发现 miR-375 可以调控多个与细胞周期、肿瘤生长和 DNA 修复相关的基因表达下调（图 7D）。而在 PC3 中，miR-375 过表达组产生了 1954 个 DEGs，其中有 1185 个上调基因和 746 个下调基因，这些基因同样涉及癌症的发生发展，以及几种生长因子信号通路和免疫(图 7C&E)。

在整合分析处于两个不同阶段细胞的 RNA-Seq 结果，我们首先发现 miR-375 在两个细胞系中调控的基因高度相似，且下调基因的重叠率要显著高于上调基因。该发现一方面验证了 RNA-Seq 结果的可靠性，另一方面也表明 miR-375 主要是通过下调其靶基因的表达量参与调控前列腺癌的发展，并且 miR-375 的调控角色会随着前列腺癌的发展而发生一定程度的转变（图 8A）。此外，我们发现在 PC3 细胞系中，miR-375 上调的基因大部分是抑癌基因，而下调的基因大部分都是促癌基因（图 8B）。另外，通过分析临床样本中 pri-PCa 和 CRPC 基因的表达情况可以发现 miR-375 大部分下调表达的基因在 pri-PCa 和 CRPC 患者体内上调表达（图 8C）。接下来，我们对受 miR-375 调控的部分癌基因和抑癌基因进行 q-PCR 验证，结果发现在两个细胞系中过表达 miR-375，其靶基因表达量变化几乎和 RNA-Seq 的结果保持一致。在 LNCap 中，miR-375 可以抑制一些"明星"癌基因 E2F1, CDK1, FOXM1 的表达，而在 PC3 中，miR-375 可以抑制 CENPF,CENPM,FOXM1 等癌基因的表达，促进能够抑制上皮间充质转化（EMT）基因 CDH1 的表达。值得注意的是，在 PC3 中 miR-375 可以促进 FOXA1 的表达而抑制 FOXA2 的表达（图 8D）。之前已有报道称，FOXA1 可以抑制 PCa 向恶性程度更高的 NEPC 转化[16]，而 FOXA2 是 NEPC 的标记基因，可以驱动 pri-PCa 向 NEPC 的转化[17]。这些结果直接从分子机制的角度证明了 miR-375 在肿瘤恶化过程中发生着抑癌的作用。总而言之，这些结果表明 miR-375 可能主要通过抑制其靶基因的表达在 PCa 中发挥抑制作用，这与之前的 miR-375 在细胞中的表型表现一致。 RNA-Seq 数据也为我们早期观察到表型，即 miR-375 抑制 PCa 细胞侵袭和转移提供了强有力的支持。例如，对 miR-375 过表达引起表达下调的基因进行功能注释，结果鉴定出这些基因功能主要涉及"转移"和"细胞周期"。此外，通过 Metscape 和 GSEA 分析结果不难发现 miR-375 下调的基因被富集在"微管细胞骨架"和"EMT"相关通路中，这说明miR-375可以抑制PCa的转移和运动能力（图 7）。同时，许多含有 KRT 家族、VIM、SNAI2、ZEB1 和 CDH1 的上皮标志物均受到 miR-375 的显著影响(图 8E)。这些结果表明，miR-375 可能增强上皮基因表达，同时诱导部分间充质基因的表达下调。

## 6  ceRNAs 竞争结合 miR-375 调控 PCa 的发展

ceRNA 理论已在各种类型的癌症中得到验证，但至今对 PCa 中 ceRNA 调控网络全景观图的认识尚不清晰，因此我们利用各种数据库，然后通过生物信息学技术的构建了

PCa ceRNA 调控网络。为了进一步验证该 ceRNA 调控网络的生物学意义，我们首先结合 LNCap 和 PC3 细胞 RNA-seq 的数据，筛选出在 miR-375 的潜在靶基因，然后将这些基因与数据库预测获得的 miR-375 靶基因进行整合分析，最后确定了 57 个基因可能是 miR-375 的直接靶点(图 9A)。在这些基因中，我们挑选了部分基因进行验证，结果发现 miR-375 可以抑制 YAP1、SEC23A、MYBL1、DPYSL3、PHTF2 在三种不同的 PCa 细胞系(LNCap、DU145 and PC3)中的表达，而 Anti-375 可以部分恢复 miR-375 介导的靶基因表达下调(图 9B&C)。上述结果证明了 miR-375 与其靶基因的负调控关系，但我们并不清楚这些靶基因是否受到 miR-375 的直接调控。为了进一步探究二者间的调控关系，我们利用 miRDB 和 starbase 数据获得了 miRNA 与靶基因的结合序列，再在此基础上构建了携带野生型或突变型目的基因的 pmir-GLOs 双荧光素酶载体。Luc 报告基因检测结果显示 miR-375 抑制了 SEC23A、MYBL1 和 PHTF2 的 3'UTR 的荧光素酶活性，这表明这三个基因受到 miR-375 的直接调控。值得注意的是，无论是 miR-375 结合位点的特异性突变还是 miR-375 的抑制剂都能恢复 miR-375 引起的荧光素酶活性降低(图 9D&E)。这些结果证实了 miR-375 与这些潜在靶基因的直接结合。

为了验证 ceRNA 网络理论存在于 PCa 中，我们选取 SEC23A 和 PHTF2 作为实验对象并将其携带荧光素基因的 3'UTR 转染到 3 个不同的细胞系（DU145、PC3 和 PC3M）中，结果发现外源 SEC23A 和 PHTF2 3'UTR 在转入细胞后，可以部分恢复 miR-375 介导的靶基因的下调,包括 CENPF, YAP1, SEC23A 和 PHTF2 等基因。值得注意的是，我们观察到外源 SEC23A 和 PHTF2 3'UTR 在不同的细胞系中可以不同程度地恢复 miR-375 靶基因的表达（图 9F-H）。这些结果说明 miR-375 和由其众多靶基因组成的 ceRNAs 是一个动态调控过程，并且这个过程会随着细胞状态和肿瘤恶性程度呈现一定差异性，即证实了我们在 PCa 中构建的 ceRNA 网络。

## 【讨论】

人类前列腺癌(PCa)是全球最常见恶性肿瘤之一，尽管当前已经有了包括手术、激素和放射治疗等治疗方法，但在原发性 PCa 患者在接受治疗后仍然面临着肿瘤复发并向 CRPC 转分化的风险。因此，我们需要挖掘 PCa 未知的发病机制，尤其是针对 CRPC 的分子机制，以确定新的治疗靶点，提高 PCa 患者的生存率。随着研究的深入，ceRNA 介导的网络调控出现在人们的视野中，即基于 ceRNAs 可以竞争结合共有的 miRNAs 而

构建的一种复杂的调控网络。ceRNA 网络可在转录后水平上解释肿瘤的发生，从而为肿瘤的诊断和治疗提供了新的指导理论以及有潜在价值的治疗策略和研究方向。但到目前为止，有关 ceRNAs 网络的报道仅涉及极少数基因的线性调控关系。例如，环状 RNA circABCC4 和 FOXP4 是一对 ceRNA 对，其中 circABCC4 可以保护 FOXP4 免受其共有的 miR-1182 的降解来促进 PCa 的进展[18]。这些 ceRNAs 和 miRNAs 可以作为前列腺癌预后的分子标志物，并对前列腺癌的恶性程度进行分层。基于该假说，我们全面分析了 399 例 pri-PCa 患者和 98 例 CRPC 样本并根据 PCa 中的 DEGs 构建了一个 ceRNA 网络。因为肿瘤生长和转移的原因很大程度上可以用 DEGs 来解释。最终，我们选择 miR-375 及其靶基因验证构建的 ceRNA 网络。值得注意的是，在我们的实验中，我们过表达 miR-375 靶基因的 3'UTR，并监控其它靶基因的表达是否受到影响，而不是关注某个特定基因，这有助于揭示和更好地理解 ceRNA 动态调控网络。

TCGA 数据库显示在 pri-PCa 患者中，CENPF 和 miR-375 均上调，这表明它们可能是 PCa 不可或缺的。如今，CENPF 已被证明是一种致癌基因，可以与 FOXM1 结合，促进前列腺癌的发展[19]。然而，在数据集 miRTarbase 中，CENPF 被鉴定为 miR-375 的靶基因之一。因此，理论上来说，CENPF 和 miR-375 不应该同时上调。这一发现表明，某些未知基因可作为"防弹背心"保护 CENPF 和 FOXM1 免受 miR-375 在 PCa 恶化中的攻击。的确，我们发现 miR-375 的两个靶基因 SEC23A 和 PHTF2，其 3' UTR 的过表达可在 PCa 各细胞系中上调 CENPF 和 FOXM1，这说明它们可能介导了 miR-375 对 CENPF 和 FOXM1 的负调控。在一定程度上该发现解释了为什么 miR-375 在 PCa 中上调却在 PCa 中发挥着抑制作用。总的来说，ceRNA 网络的构建是成功的。其中，miR-375 作为 ceRNA 网络中的枢纽之一，为我们证实了 ceRNA 网络在 PCa 中的存在。我们的研究将增加我们对 ceRNA 网络在 PCa 预后作用的理解。此外，参与 ceRNA 网络构建的候选基因可以进一步作为前列腺癌的潜在治疗靶点和预后生物标志物进行系统地分析与评估。

## 【材料与方法】

### 1　ceRNA 网络的构建和网络核心 RNA 的鉴定

我们通过四个数据库揭示了 DE miRNAs 和 DE mRNAs 之间的相互作用，它们分别是 TargetScan (http://www.targetscan.org/vert_80/),miRTarBase

(https://maayanlab.cloud/Harmonizome/resource/MiRTarBase),starbase(https://starbase.sysu.edu.cn/starbase2/), miRDB (http://mirdb.org/)。然后，利用基于高通量表达数据的生物分子相互作用网络连接可视化软件 Cytoscape 构建 PCa 中的 ceRNA 网络景观。s

## 2 细胞培养

DU145、PC3 和 LNCap 细胞购自中国科学院上海细胞库。培养方式：在 RPMI 1640 培养基(Gibco, 赛默飞，北京，中国)中添加 10%胎牛血清(FBS) (GEMINI, 卡拉巴萨斯, 加利福尼亚州，美国)，1%青霉素-链霉素(100×) (新赛美生物科技有限公司，苏州，浙江，中国)，并于 37°C 加 5% CO2 的湿空气中培养。

## 3 瞬时转染 miRNA mimics 和 inhibitor

我们使用 Lipofectamine 2000 (Invitrogen, Life Technologies, 卡尔斯巴, 加利福利亚州, 美国)在 DU145、PC3 和 LNCap 细胞中转染 50-100nM 的非靶向 NC oligos 或 miRNA mimics，而 NC inhibitor 或 miRNA inhibitor 使用剂量为 100-200nM 之间。所用 mimics 和 inhibitor 购买于吉玛公司(苏州，浙江，中国)，转染过程根据制造商提供的说明书进行。培养 24-72h 后，利用转染细胞进行表型实验或分子实验。

文章中所用 miRNA mimics 和 inhibitor 序列：

| 序列 | 有义链 | 反义链 |
| --- | --- | --- |
| NC oligos | 5'-UUCUCCGAACGUGUCACGUTT-3' | 5'-ACGUGACACGUUCGGAGAATT-3' |
| miR-375 mimics | 5'-UUUGUUCGUUCGGCUCGCGGA-3' | 5'-ACGCGAGCCGAACGAACAAAUU-3' |
| NC inhibitor | 5'-CAGUACUUUUGUGUAGUACAA-3' | |
| miR-375 inhibitor | 5'-UCACGCGAGCCGAACGAACAAA-3' | |

## 4 RNA 提取及 q-PCR 分析

miRNeasy Tissue/ cells Advanced Mini Kit (QIAGEN, 杜塞尔多夫, 德国) 被用于从培养细胞中提取总 RNA。miRNA 的表达水平用 Harpin-itTM microRNA 和 U6 snRNA 标准化 RT-PCR 定量试剂盒(吉玛公司)进行检测，mRNA 逆转录过程采用 HiScript II Q RT SuperMix for qPCR (+gDNA wiper) 试剂盒(诺唯赞，南京，江苏，中国)，表达水平采用 2× SYBR Green qPCR Master Mix 试剂 (Biomarker，北京，中国)和 CFX96 Touch Real-time PCR 检测系统(Bio-rad, 伯克利, 加利福利亚州, 美国)进行实时 q-PCR 检测。q-PCR

结果分析采用 Delta-Delta-Ct (DDCt)法，该方法可计算差异基因表达的倍数变化。上述实验的具体过程以公司提供的说明书为准。

文章用于检测 miR-375 表达及其靶基因表达的引物序列（物种：人类）：

| miR-375 引物序列 | 序列 |
| --- | --- |
| miR-375-FO | TGCTTTGTTCGTTCGGCTC |
| miR-375-RE | TATGGTTGTTCACGACTCCTTCAC |
| miR-375-PR | CCAACCATACAGACTCACGCGA |

| miR-375 靶基因引物序列 | 序列 |
| --- | --- |
| SEC23A-F | TGCTAGGAACTGGGCAGATG |
| SEC23A-R | AGCTGCCTCCTGGTCAAAAG |
| MYBL1-F | GAAATCGTTGGGCAGAAATTGC |
| MYBL1-R | TGACACATACTGATACCCAGGG |
| DPYSL3-F | GACCGTCTCCTTATCAAGGGA |
| DPYSL3-R | GCATCTGGAAGTGAGTATGGAC |
| PHTF2-F | AACACCCAAACCTCCTCTAAGT |
| PHTF2-R | GTGCTTGTACCGTGGTTCTG |
| YAP1-F | TCCTTAACAGTGGCACCTATCAC |
| YAP1-R | TCACCTGTATCCATCTCATCCA |
| hCDK1-F | TACACATGAGGTAGTAACACTCTG |
| hCDK1-R | AGTCCTGTAAAGATTCCACTTCTG |
| FOXA1-F | GGGGGTTTGTCTGGCATAGC |
| FOXA1-R | GCACTGGGGGAAAGGTTGTG |
| FOXA2-F | GGAGCAGCTACTATGCAGAGC |
| FOXA2-R | CGTGTTCATGCCGTTCATCC |
| FOXM1-F | CGTCGGCCACTGATTCTCAAA |
| FOXM1-R | GGCAGGGGATCTCTTAGGTTC |
| CDH1-F | GGGGTCTTGCTATGTTGCC |
| CDH1-R | CAATACCTAGTCAAGATGTGGC |
| CENPF-F | AAAGAAACAGACGGAACAACTG |
| CENPF-R | CCAAGCAAAGACCGAGAACT |
| CENPM-F | CCAGAACACAGAGGAGTC |
| CENPM-R | CAGGTCACAGTAGAGCAG |
| EXO1-F | CACCATGGGGATACAGG |
| EXO2-R | TTACTGGAATATTGCTCTTTG |

**5 细胞增殖**

通过人工细胞计数的方式和克隆形成试验观察细胞增殖能力。对于细胞计数实验，我们将大约 20k 转染过的 DU145、PC3 和 PC3M 铺到 12 孔板中，12K LNCap 细胞铺到 24 孔板中。铺板后于 37℃下孵育过夜以贴壁，12-24 小时后测量贴壁细胞的数量以确定细胞数量真实起始值，并而后每隔一天对细胞进行计数，并将数量的变化以折线图的方式绘制生长曲线。为了进行克隆形成试验，我们先将细胞株种植到 6 孔板中。值得注意的是，细胞铺板数量会根据每种细胞的生长特性进行调整，对于 DU145、PC3 和 PC3M 来说，我们将数量控制在 6k-8k 之间，而对于 LNCap 来说，数量将被控制在 20k-30k 之间。在 37℃下培养约 9 天后，从培养箱取出细胞，首先除去培养基并用磷酸盐缓冲盐水（PBS）清理漂浮于 6 孔板的细胞碎片，而后用 4%多聚甲醛固定克隆板 15 分钟。接下来除去固定液并再次用 PBS 轻轻清洗克隆板，然后在室温下用 0.1%结晶紫染色 15 分钟。最后扫描克隆板并利用倒置荧光显微镜(MshOt，广州，广东，中国)对克隆进行拍照，然后计算出照片灰度值以进行定量分析。

**6 细胞凋亡分析**

细胞凋亡检测采用 APC-Annexin V/PI 细胞凋亡试剂盒(US EVERBRIGHT，上海，中国)。在 DU145、PC3 细胞转染 NC 核苷酸链或者 miR-375 mimics 后的第 5 天进行 Annexin 和 PI 染色，然后在流式仪 CytoFLEX (BECKMAN COULTER, Life Sciences, 上海，中国)中进行检测荧光，并使用以 CytExpert 命名的配套软件进行分析数据。实验过程参照公司提供说明书

**7 伤口愈合试验**

将大约 400k 转染 NC 核苷酸链（或抑制剂）或者 miR-375 mimics（或抑制剂）的 DU145、PC3 细胞铺入 6 孔板中。当平板上约 90%覆盖细胞时，我们用消过毒的钝型枪头轻轻地划直线，然后用 PBS 冲洗两次。最后，用含少量血清的胎牛血清（3%-5% FBS）的 RPMI 1640 培养基培养细胞，并在倒置荧光显微镜下分别观察 0、12 和 24 小时创面愈合率。

## 8 肿瘤成球实验

将转染 48 小时的 DU145 和 PC3 单细胞悬液(约 2k 细胞/ml)加入超低附 6 孔板,并在添加 50 × B27、20 ng/ml 碱性成纤维细胞生长因子(bFGF)和 40 ng/ml 表皮生长因子(EGF)的干细胞培养基中培养。约 2 周后,在倒置荧光显微镜下观察肿瘤球的大小和数量并记录。

## 9 Transwell 细胞迁移

先在 24 孔板的下室中加入约 600-700μL 20% FBS 培养液,后将上室加入 24 孔板,然后在上室中加入 30-50k/well 细胞悬液(细胞悬液不含或含 1% FBS 培养液,且悬液体积不超过 200μL)。24-48h 后,取出 24 孔板于普通相差光学显微镜下观察,如若下室中开始出现细胞则可停止培养,然后弃上室内的培养液,并加入 600-700μL 4%PFA 固定 10min,再用 0.1%结晶紫染液对小室进行染色。20min 后回收结晶紫染液,清洗后于显微镜下拍 9 张照片进行灰度值分析。

## 10 皮下瘤模型构建

我们采用 PC3 和 PC3M 细胞来构建小鼠皮下瘤模型:先将 6-8 周龄的 BALB/c 小鼠随机分为对照组(NC)和实验组(miR-375 过表达),每组小鼠设置 3-4 只。将转染的 PCa 细胞消化离心处理成细胞悬液状态,然后在显微镜下计数并将细胞悬液浓度调整至 $1\times10^6$ 个/50μL,再将细胞悬液和基质胶按照 1:1 的比例注射于小鼠左、右侧腋下的皮下部位。注射后每隔两天测一次小鼠肿瘤大小,肿瘤体积=(长 x 宽 2)/2。待两个月的时间(小鼠肿瘤体积达 1000mm3 左右时)通过 $CO_2$ 安乐死法处理小鼠,然后解剖肿瘤,绘制小鼠肿瘤生长曲线。

## 11 肺转移实验

肺转移模型是当下用于研究体内肿瘤转移最常见的方法之一。在本课题中,我们采用 PC3M 细胞来构建小鼠肺转移模型:先将 6-8 周龄的 BALB/c 雄鼠随机分为对照组(NC)和实验组(miR-375 过表达),每组小鼠设置 5 只。将 PCa 细胞消化离心处理成细胞悬液状态,然后在显微镜下计数并将细胞悬液浓度调整至 $1\times10^6$ 个/100μL 后,通过尾静脉注射入小鼠体内,一段时间后每隔两天观察小鼠状态,当小鼠状态变差时,例如行动

迟缓、体重减轻等，通过脱颈法处死小鼠，解剖取出肺部组织，观察肿瘤在肺部的转移情况并进行数据统计。

## 12  荧光素酶报告试验

通过 PCR 扩增了 miR-375 靶基因 SEC23A、PHTF2 和 MYBL1 的野生型和突变型 3 ' UTR 部分序列。将这些与 miR-375 具有潜在结合位点的序列各自嵌入 pmirGLO 双荧光素酶 miRNA 靶表达载体(麦迪逊，威斯康星州，美国)中，然后将其转染到 NC 或 miRNA mimics 处理的细胞。48h 后用同公司提供的蛋白核酸分析仪对荧光素酶活性进行检测。

用于构建 SEC23A、PHTF2 和 MYBL1 野生型和突变型 3 ' UTR 的引物序列：

| 基因 | 引物编号 | 引物序列(5'->3') | 引物功能 | 扩增片段大小 | GeneBank 编号 |
|---|---|---|---|---|---|
| MYBL1 | MYBL1-F-A | ctcgctagcctcgagtctagaTAAAGTTGTAAGATAGCCC | 扩增 | 243bp | NC_000008.11 |
| | MYBL1-R-A | cttgcatgcctgcaggtcgacACTGCAGCATCAAGAGAGT | | | |
| | MYBL1-F-M | gattttaaaaattgttttaaaAATAATgatgggaaaataatagaatg | 突变 | | |
| | MYBL1-R-M | cattctattattttcccatcATTATTtttaaaaacaattttaaaaatc | | | |
| SEC23A | SEC23A-F-A | ctcgctagcctcgagtctagaCCATTTATCTGTGGAAAC | 扩增 | 232bp | NC_006364 |
| | SEC23A-R-A | cttgcatgcctgcaggtcgacACATCTGTAGTACGAGGC | | | |
| | SEC23A-F-M | gaattataatgagagcaataaAATAATTtttattttgcttaccacag | 突变 | | |
| | SEC23A-R-M | ctgtggtaagcaaaataaaAATTATTttattgctctcattataattc | | | |
| PHTF2 | PHTF2-F-A | ctcgctagcctcgagtctagaAACTTGCCTTACTTGAGG | 扩增 | 218bp | NC_000007.14 |
| | PHTF2-R-A | cttgcatgcctgcaggtcgacAAAGAGCACTTGTTGACC | | | |
| | PHTF2-F-M | gtttctcttgaattattttg AATAAT tgccaggatccaaactg | 突变 | | |
| | PHTF2-R-M | cagtttggatcctggca ATTATT caaaataattcaagagaaac | | | |

注意：A 代表扩增引物序列； M 代表突变引物序列

## 13  转录组测序技术（RNA-Seq）

先用 RNeasy®Mini Kit (QIAGEN)提取总 RNA（总量>1ug），然后构建测序文库并由 LC-Bio 科技有限公司(杭州，浙江，中国)进行高通量测序。值得注意的是，在 RNeasy®Mini Kit 中我们使用了 DNA 酶试剂以避免 DNA 对测序结果产生影响。在获得初始数据后，我们首先使用 R 包 DESeq2 对 DEGs 的表达进行了分析[20]。然后利用基因集富集分析(Gene set enrichment analysis, GSEA)和 Metscape(http://metascape.org)对 DEGs 进行了功能注释。GSEA 是一种基于基因集的富集分析方法，在对基因表达数据分析时，首先确定分析的目的，即选择 MSigDB 中的一个或多个功能基因集进行分析，然后基于基因表达数据与表型关联度的大小进行排序。最后通过判断每个基因集内的基因是否富集于表型相关度排序后基因列表的上部或下部来判断此基因集内基因的变化对表

型变化的影响 [21,22]。Metascape 是一个可靠而强大的功能富集分析工具，集成了来自 40 多个知识库的注释信息，可全面分析基因的功能，并将结果以柱状图的形式可视化 [23]。最后，用 R 包 heatmap3 展示 DEGs，bubble 包展示 Metascape 结果。本文章中所有 RNA-Seq 数据已存入 NCBI Gene Expression Omnibus，所有其他相关数据可根据要求从相应作者处获得。

## 14　统计分析

一般来说，在可行的情况下，每种条件都做了三次实验。结果使用 GraphPad Prism 8 计算，并以均数±标准差表示。大多数采用非配对双尾 t 检验来确定统计学差异，部分采用双因素方差分析。

【图与图注】

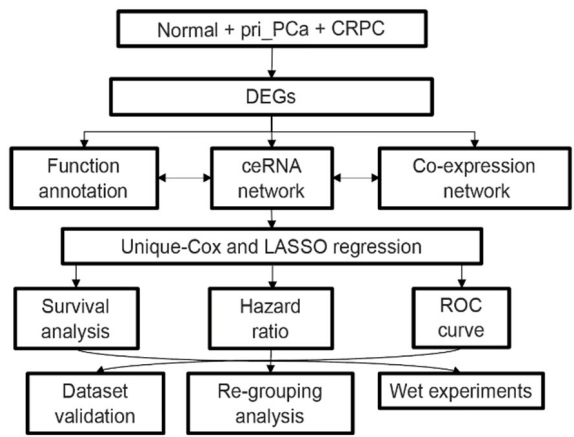

**图 1 ceRNA 网络构建的研究框架**

**Fig.1 Research framework for ceRNA network construction**

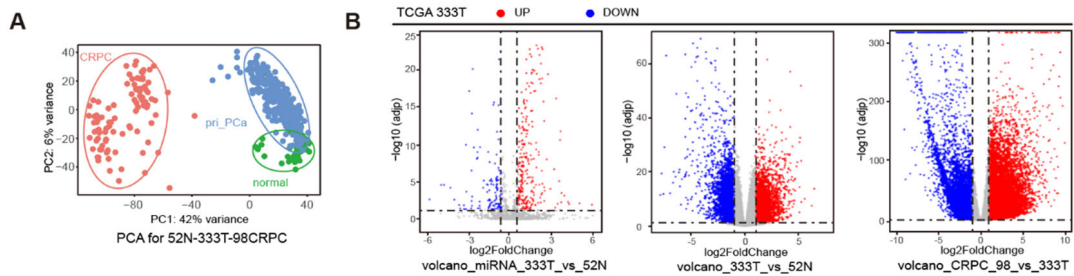

**图 2 PCa 各阶段差异表达基因**

**Fig.2 The DEGs in different stages of PCa**

A：基于 PCa 各阶段差异表达基因将前列腺癌分为三类; B：PCa 各阶段差异表达基因火山图。红色和蓝色的点分别代表上调和下调的基因;黑色的点代表变化不显著或变化倍数小的基因。

A: PCa is classified into three group based on DEGs in different stages of PCa; B: The volcano map of DEGs in different stages of prostate cancer. The red and blue dots represent up-and down-regulated genes,respectively,and black dots the RNAs that do not reach the threshold for screening differentially expressed genes

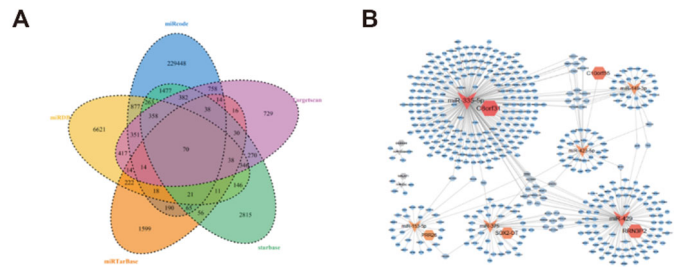

**图 3 PCa 中的 ceRNAs 调控网络**

**Fig.3 ceRNAs network in PCa**

A：基于 miRTarBase, starbase, miRDB, TargetScan 四个数据库初步筛选 DE ceRNAs 及与其具有潜在结合位点的 miRNAs；B：PCa 中的 ceRNAs 调控网络。

A: DE ceRNAs and miRNAs with common potential binding sites were preliminarily screened Based on miRTarBase, starbase, miRDB and TargetScan; B: ceRNAs network in PCa.

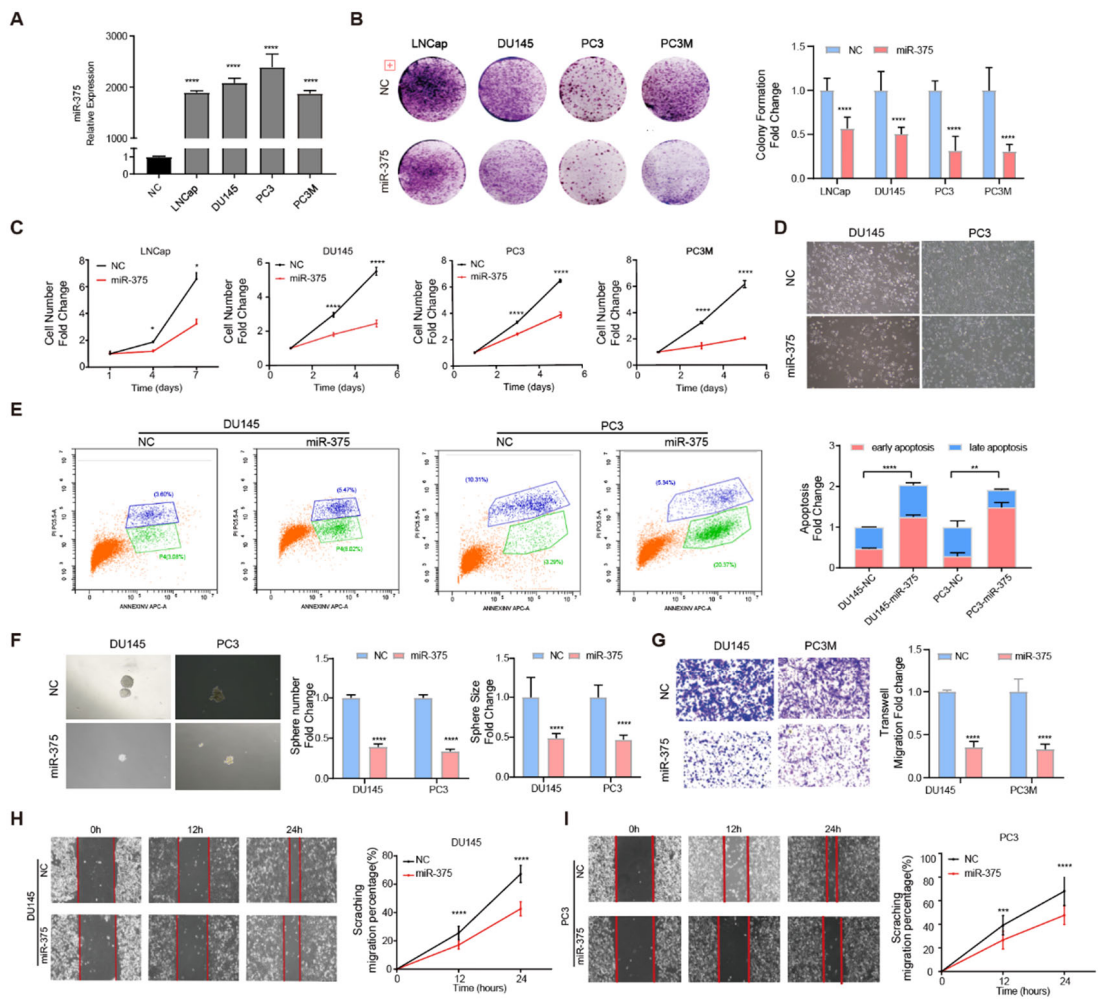

**图 4 miR-375 过表达能够抑制前列腺癌的进展**

**Fig.4 miR-375 over-expression enables to inhibit the progression of prostate cancer**

A：与相应的阴性细胞相比，转染 miR-375 mimics 的 4 种 PCa 细胞系(LNCap, DU145, PC3 和 PC3M) 中 miR-375 的相对表达水平。数据经双尾非配对 t 检验，并以均值±s.d. ****p<0.0001 展示；B&C：转染 NC oligos 或 miR-375 mimics 的 DU145 和 PC3 细胞增殖能力通过克隆形成试验(展示有代表性的孔，B，左)和显微细胞计数试验(C)进行评估，并通过非配对 t 检验进行统计分析(B，右，C)，*p<0.05，****p<0.0001；D：NC oligos 或 miR-375 过表达对细胞形态的影响(DU145 和 PC3)；E：流式细胞术检测(展示有代表性图像)分析 NC oligos 或 miR-375 mimics 诱导 DU145 和 PC3 凋亡水平。数据以均值±s.d 表示。****p<0.0001，**p<0.01，双尾未配对 t 检验；F：miR-375 过表达抑制 DU145 和 PC3 成球能力。P 值的计算采用非配对 t 检验面板(F，中和右)，****p<0.0001；G-I：显示具有代表性的小室和划痕实验图像(G-I，左)，并对转染 NC 寡核苷酸或 miR-375 mimics 的 DU145 和 PC3 的侵袭和迁移能力进行了定量。***<0.001，****p<0.0001，双尾未配对 t 检验分析(G-

I，右)。

A: Relative expression levels of miR-375 in 4 PCa cell lines (LNCap, DU145, PC3 and PC3M) transfected with miR-375 mimics compared with the corresponding negative cells. Data are represented as means ± s.d. ****p <0.0001, two-tailed unpaired t-test; B&C: The proliferation ability of DU145 and PC3 cells transfected with NC oligos or miR-375 mimics was evaluated by colony formation assay (representative wells were presented, B, left) and microscopic cell counting assay(C) with statistical analysis by unpaired t-test (B, right and C), *p<0.05，****p<0.0001; D: The effect of NC oligos or miR-375 over-expression on cellular morphology (DU145 and PC3); E: Flow cytometry assay (representative images were presented) was used to analyze the apoptosis levels of DU145 and PC3 induced by NC oligos or miR-375 mimics. Data are represented as means ±s.d. ****p <0.0001, **p<0.01, two-tailed unpaired t-test; F: miR-375 over-expression inhibited tumor sphere formation ability of DU145 and PC3. P values were calculated using unpaired t-test for panels (F, middle and right), ****p<0.0001; G-I: Representative trans-well and scratching assay images were shown (G-I, Left) and quantification of invasion and migration ability formed by DU145 and PC3 transfected with NC oligos or miR-375 mimics was plotted. ***<0.001, ****p <0.0001 by two-tailed unpaired t-test analysis (G-I, Right).

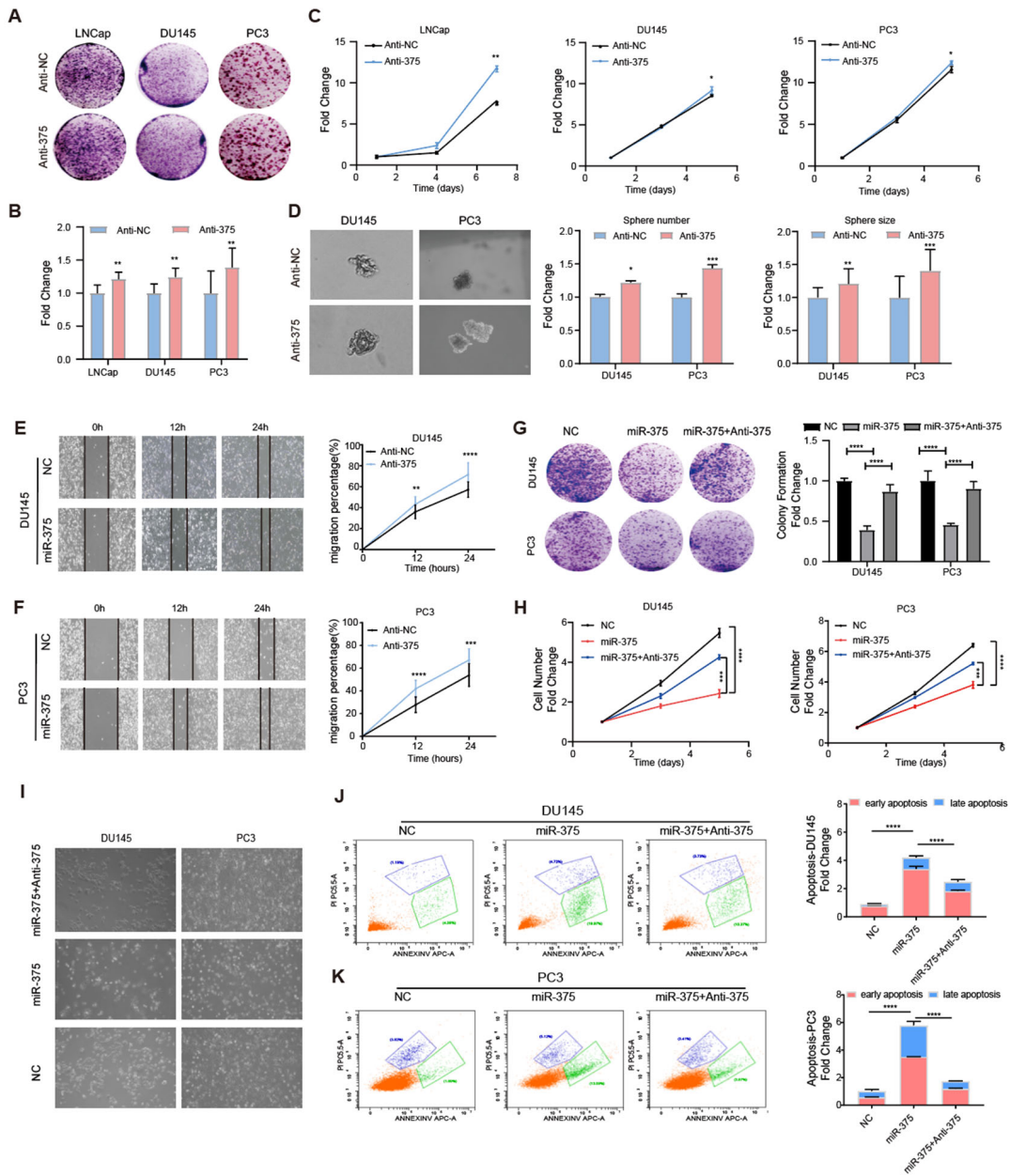

**图 5 miR-375 抑制剂前列腺癌的发展无明显的影响**

**Fig.5 miR-375 inhibitor has no significant effect on the progression of prostate cancer**

A-C：转染 NC inhibitor 或 miR-375 inhibitor 的 DU145 和 PC3 细胞存活率通过克隆形成试验(展示有代表性的孔，A)和显微细胞计数试验(C)进行评估，并通过非配对 t 检验进行统计分析(B,C), *p<0.05, **p<0.01；D：NC inhibitor 或 miR-375 inhibitor 对 DU145 和 PC3 成球能力的影响。P 值的计算采用非配对 t 检验面板(D，中和右), *p<0.05，**p<0.01，***p<0.001；E&F：展示具有代表性的划痕实验图像(E,F，左)，并对转染 NC 抑制剂或 miR-375 抑制剂的 DU145 和 PC3 的迁移能力进行了定量。**p<0.01，***p<0.001，

****p<0.0001,双尾未配对 t 检验分析(E,F，右)；G：分别用 NC oligo，miR-375 mimics 或 miR-375 mimics 加 miR-375 inhibitor 孵育 DU145 和 PC3 细胞 48 小时后，通过克隆形成试验评估细胞活性(展示具有代表性的图片，G，左图)，并通过未配对 t 检验进行统计分析，****<0.0001 (G，右图)；H：通过显微细胞计数试验评估细胞活性，并通过双因素方差分析进行统计分析，***p<0.001, ****p<0.0001；I：NC oligos 或 miR-375 过表达对 DU145 和 PC3 形态的影响；J&K： 通过流式细胞仪(展示有代表性图像)观察 NC oligos 或 miR-375mimics 诱导 DU145(J)和 PC3(K)凋亡情况，并进行 PCa 细胞凋亡水平的定量分析 DU145(J，右图)和 PC3(K，右图)。数据以均值±s.d 表示。****p<0.0001，**p<0.01，双尾未配对 t 检验。

A-C: The survival of DU145 and PC3 cells transfected with NC inhibitors or miR-375 inhibitors was assessed by clonal formation assay (representative pore shown, A) and microscopic cell count assay (C), and statistically analyzed by unpaired T-test (B,C), *p<0.05, **p<0.01; D: The effect of NC inhibitors or miR-375 inhibitors on pellet formation ability of DU145 and PC3. P-values were calculated using the unpaired T-test panel (D, middle and right),*p<0.05，**p<0.01，***p<0.001; E&F: representative scratch experimental images (E,F, left) was displayed and quantification of migration ability formed by DU145 and PC3 transfected with NC oligos or miR-375 mimics was plotted. **p<0.01，***p<0.001，****p <0.0001 by two-tailed unpaired t-test analysis (E,F, right); G: DU145 and PC3 cells were incubated with NC oligos, miR-375 mimics or miR-375 mimics added with miR-375 inhibitor respectively for 48 h. Cell viability was assessed by clonal formation assay (representative wells were presented, G, left panel) with statistical analysis by unpaired t-test, ****p <0.0001 (G, right panel). H: Cell viability was assessed by microscopic cell counting assay with statistical analysis by unpaired t-test, ***p<0.001, ****p <0.0001 (H, right panel).I: The effect of NC oligos or miR-375 over-expression on DU145 and PC3 cellular morphology; J&K: Flow cytometry assay (representative images were presented) was used to observe the apoptosis levels of DU145(J) and PC3 (K) induced by NC oligos or miR-375 mimics. Quantitative analysis of DU145(J, right panel) and PC3 (K, right panel) cell apoptosis. Data are represented as means ±s.d. ****p <0.0001, **p<0.0, two-tailed unpaired t-test.

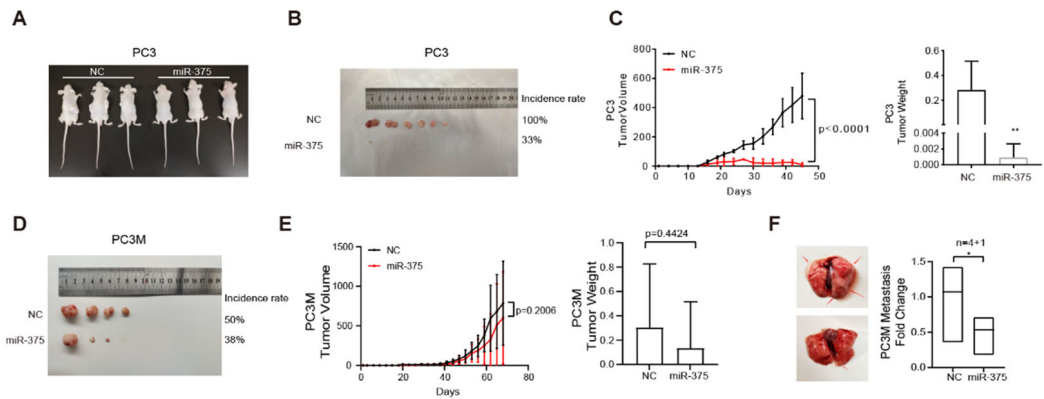

**图 6 miR-375 在体内过表达可抑制前列腺癌的发展**

**Fig.6 miR-375 over-expression supresses the progression of PCa in vivo**

A,B：展示了植入 NC 或 miR-375 过表达 PC3 细胞的 6 只 BALB/c 裸鼠(A)和肿瘤图像(B)；C：植入后指定天数内 PC3 异种移植瘤体积(C，左)和重量(C，右)的定量。n = 3 只(NC)或 3 只(miR-375)小鼠。数据以均值±s.d. ****p<0.0001 表示，双因素方差分析 (miR-375 与 NC，C, 左); **p = 0.0037，双尾未配对 t 检验分析 (miR-375 与 NC, C, 右)；D：展示了 8 个来自 NC 或 miR-375 过表达 PC3M 细胞的异种移植瘤图像；E：植入后指定天数内 PC3M 异种移植瘤体积(E，左) 和重量(E，右) 的定量。n = 4 只(NC)或 4 只(miR-375) 小鼠。数据以均数±s.d. p=0.2006，双向方差分析 (miR-375 与 NC, E，左); p = 0.4424，双尾非配对 t 检验分析 (miR-375 与 NC, E，右)；F：肺表面可见结节图像。将转染 NC 寡核苷酸或 miR-375 mimics 的 PC3M 细胞注入裸鼠尾静脉 (每组 n =5)，2 个月后处死小鼠。数据以均数±s.d 表示，*p=0.0311，双尾未配对 t 检验。

A,B: 6 BALB/c-Nude mice (A) implanted by PC3 cells expressing NC or miR-375 and tumor images(B) were presented; C: The quantification of PC3 xenografts tumor volume (C, left) and weight (C, right) on the indicated days after implantation. n = 3 (NC) or 3 (miR-375) mice. Data were presented as means ±s.d. ****p <0.0001, two-way ANOVA analysis (miR-375 versus NC, C, left); **p = 0.0037, two-tailed unpaired t-test analysis (miR-375 versus NC, C, right); D: The tumor image of 8 xenografts derived from NC or miR-375 over-expression PC3M cells was presented; E: The quantification of PC3M xenografts tumor volume (E, left) and weight (E, right) on the indicated days after implantation. n = 4 (NC) or 4 (miR-375) mice. Data were presented as means ±s.d. p=0.2006, two-way ANOVA analysis (miR-375 versus NC, E, left); p = 0.4424, two-tailed unpaired t-test analysis (miR-375 versus NC, E, right); F: Images of visible nodules on the lung surface. PC3M cells transfected with NC oligos or miR-375 mimics were injected into each nude mouse tail vein (n =5 for each group), and the mice were sacrificed 2 months later. Data are shown as means ± s.d. *p=0.0311, two-tailed unpaired t-test.

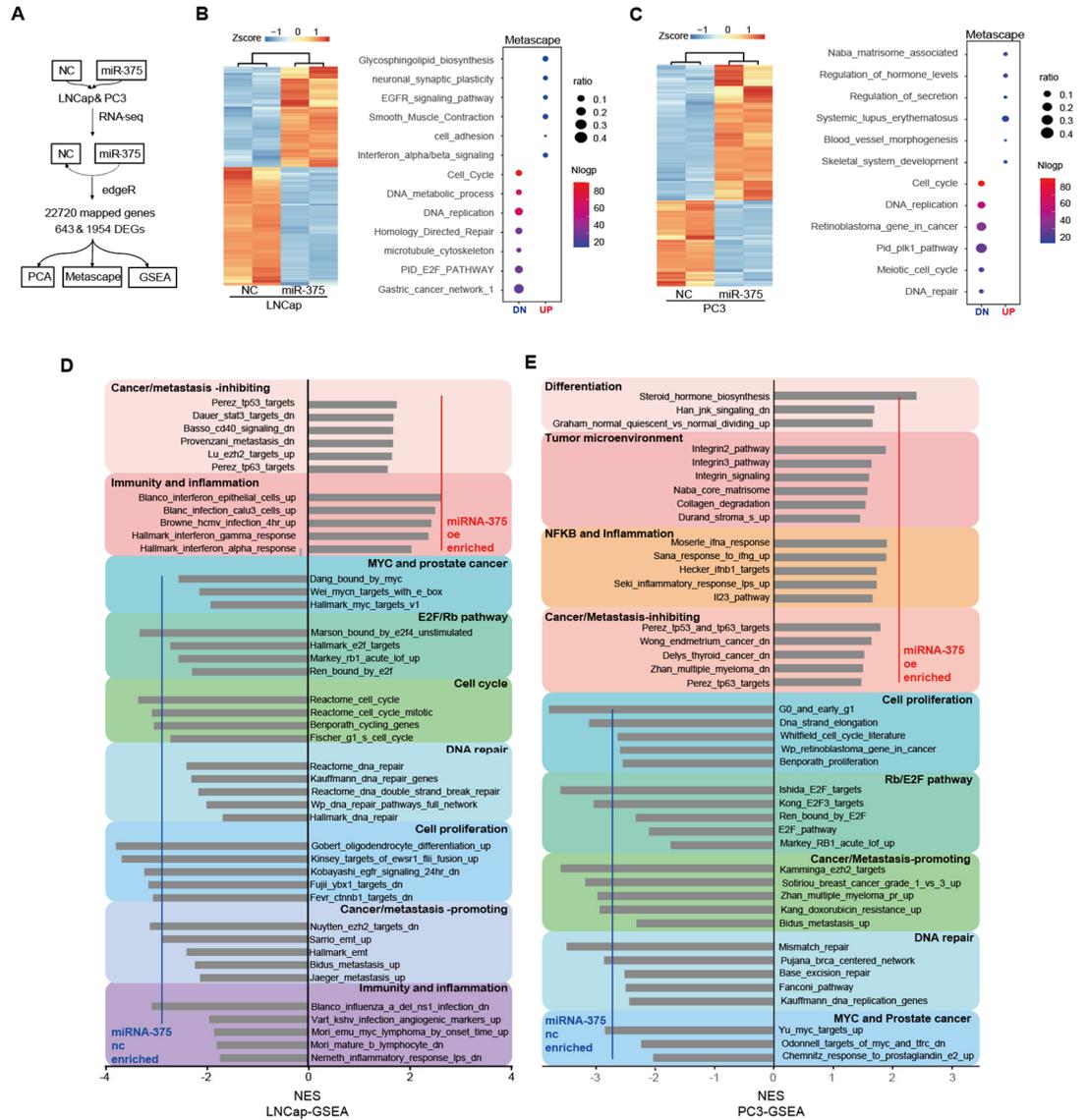

**图 7 RNA-Seq 分析揭示了新的 miR-375 靶向信号分子和通路**

**Fig.7 RNA-Seq analysis reveals novel miR-375 targeted signaling molecules and pathways**

A：LNCap 和 PC3 细胞 RNA-Seq 的实验方案及后续数据分析；B&D：综合分析 LNCap 细胞 RNA-Seq 结果，包括以热图形式展示的 DEGs（B，左），Metascape 的功能注释(B，右)和 GSEA 揭示的基因富集情况(D)；C&E：综合分析 PC3 细胞 RNA-Seq 结果，包括以热

图形式展示的 DEGs (C，左)， Metascape 的功能注释(C，右)，GSEA 揭示的基因富集情况(E)。

A: Experimental scheme of RNA-Seq in LNCap and PC3 cells and subsequent data analysis; B&D: The comprehensive analysis of LNCap RNA-Seq result, including heat map of DEGs (B, left), functional annotations through Metascape(B, right) and genes enrichment revealed by GSEA(D); C&E: The comprehensive analysis of PC3 RNA-Seq result, including heat map of DEGs (C), functional annotations through Metascape(E, left) and genes enrichment revealed by GSEA(E, right).

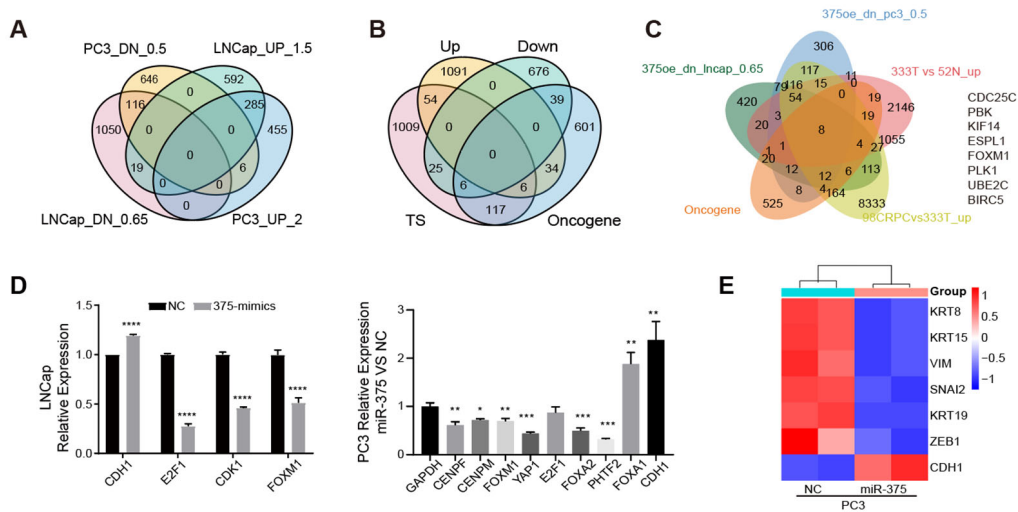

**图 8 miR-375 在 PCa 中发挥抑癌功能的分子机制**

**Figure 8 The molecular mechanism of miR-375 acting as a tumor suppressor in PCa**

A：LNCap 和 PC3 RNA-Seq 结果中 DEGs 的重叠图；B：PC3 RNA-Seq 中的 DEGs 与致癌基因和抑癌基因的重叠图；C：临床 pri-PCa 和 CRPC 中上调基因与 LNCap 和 PC3 中 miR-375 下调基因的重叠图像；D：通过 q-PCR 在 LNCap（D,左图）和 PC3(D,右图)检测部分 miR-375 候选靶基因的表达；所有数据经双尾非配对 t 检验，并以均值±s.d 表示，其中 *p<0.05，**p<0.01，***p<0.0001，<****p<0.0001 展示；E：PC3 RNA-Seq 中部分转移标记物的表达变化。

A: Overlapping image of DEGs identified from LNCap and PC3 RNA-Seq;B: Overlapping image of DEGs derived from PC3 RNA-Seq with oncogenes and tumor suppressor; C: Overlapping image of up-regulated genes in clinical pri-PCa and CRPC with miR-375-induced down-regulated genes in LNCap and PC3; D: Partial miR-375 candidate target genes expression in LNCap (D, left panel) and PC3 (D, right panel) followed by q-PCR; Data was analyzed by unpaired two-tailed t-test and presented as mean ±s.d, *p<0.05, **p<0.01, ***p<0.0001, <****p<0.0001; E: The expression alteration of partial metastasis markers in PC3 RNA-Seq.

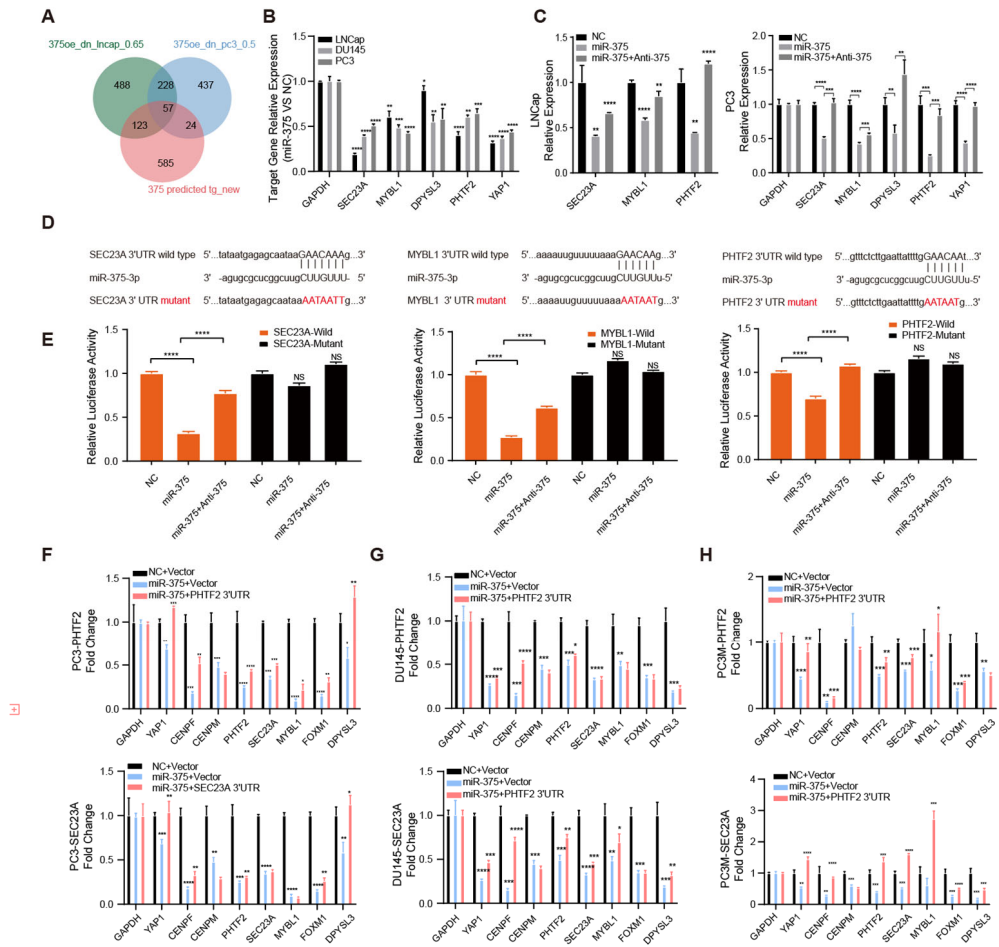

**图 9 在 PCa 细胞系中的验证 ceRNA 网络**

**Figure.9 ceRNA network validation in PCa cell lines**

A：根据 RNA-Seq 数据和相关数据库筛选 miR-375 的候选靶基因；B：通过 q-PCR 检测转染 NC oligos 或 miR-375 mimics 的 PCa 细胞系(LNCap, DU145 和 PC3)中 miR-375 靶基因的相对表达量。在图中已显示平均值±s.d.值；C：通过 q-PCR 检测在 NC oligos、miR-375 mimics 或 miR-375 抑制剂的 miR-375 mimics 和 inhibitor 共转染过的 LNCap(C，左图)和 PC3(C，右图)细胞中 miR-375 靶基因的相对表达量。在图中已显示平均值±s.d.值；D：miR-375 与 3 个潜在靶基因 3'UTR, 包括 SEC23A(D，左)、MYBL1(D，中)、PHTF2(D，右)之间的潜在结合位点和相关突变序列；E：荧光素酶报告基因/诱变试验表明，miR-375 抑制 SEC23A (E，左)、MYBL1(E，中)、PHTF2(E，右) 3'UTR 荧光素酶的活性。通过 miR-375 抑制剂阻断 miR-375 的功能或者突变结合位点，都可使得荧光素酶活性得以恢复。在图中已显示平均值±s.d.值，****p<0.0001;F-H::通过动态调控 miR-375 与靶基因之间在 3 个 PCa

细胞系(PC3,F;DU145 G;PC3M,H)的表达来验证 ceRNA 网络。在图中已显示平均值±s.d.值。

A: Screening candidate target genes of miR-375 according to RNA-Seq data and bioinformatic analysis; B: Relative expression of target genes of miR-375 revealed by q-PCR in 3 PCa cell lines transfected with NC oligos or miR-375 mimics (LNCap, DU145 and PC3). Mean ± s.d. values are shown; C: Relative expression of target genes of miR-375 revealed by q-PCR in LNCap (C, left) and PC3(C, right) transfected with NC oligos, miR-375 mimics or miR-375 mimics added with miR-375 inhibitor. Mean ± s.d. values are shown; D: Predicted binding sites and relevant mutation sequences between miR-375 and 3 potential target genes 3' UTR including SEC23A (D, left), MYBL1(D, middle), PHTF2(D, right); E: Luciferase reporter / mutagenesis assay indicated that luciferase activity of SEC23A (E, left), MYBL1(E, middle), PHTF2(E, right) 3'UTR was repressed by miR-375. Luciferase activity was rescued due to the function of miR-375 was blocked by miR-375 inhibitor or the mutation of binding sites. ****$p<0.0001$, mean ± s.d. values are shown; F-H: ceRNA network validation in 3 PCa cell lines(PC3,F; DU145,G; PC3M,H) through dynamic regulation the expression between miR-375 and target genes. Mean ± s.d. values are shown.